\def\noeditingmarks{1}
\documentclass[twocolumn]{article}
\usepackage{fullpage}
\usepackage[utf8]{inputenc}
\usepackage[usenames,dvipsnames]{xcolor}
\usepackage{url}
\usepackage{xspace}
\usepackage{graphicx}
\usepackage[font=small,skip=0pt]{caption}
\usepackage{times}
\usepackage[small,compact]{titlesec}
\usepackage{enumitem}

\newcommand\rmv[1]{}

\def\helvsize{\small}
\def\hn{\usefont{OT1}{phv}{mc}{n}\selectfont\helvsize}

\definecolor{orange}{rgb}{1,0.5,0}

\newcommand{\mpfont}{\hn\scriptsize}
\newcommand{\MPworker}[2]{{\color{#1}\vrule\vrule}{\marginpar{\color{#1}\mpfont #2}}}
\if\noeditingmarks 1
  \newcommand{\MPmka}[1]{}
  \newcommand{\MPkk}[1]{}
  \newcommand{\MPss}[1]{}
  \newcommand{\MPsn}[1]{}
  \newcommand{\kim}[1]{}
  \newcommand{\marcos}[1]{}
  \newcommand{\sharad}[1]{}
  \newcommand{\stanko}[1]{}
  \newcommand{\todo}[1]{}
  \newcommand{\TODO}[1]{}
\else
  \newcommand{\MPmka}[1]{\MPworker{blue}{#1}}
  \newcommand{\MPkk}[1]{\MPworker{orange}{#1}}
  \newcommand{\MPss}[1]{\MPworker{teal}{#1}}
  \newcommand{\MPsn}[1]{\MPworker{violet}{#1}}
  \newcommand{\kim}[1]{\noindent{\color{Blue} {\bf \fbox{KK} {\it#1}}}}
  \newcommand{\marcos}[1]{\noindent{\color{Red} {\bf \fbox{MA} {\it#1}}}}
  \newcommand{\sharad}[1]{\noindent{\color{Green} {\bf \fbox{SS} {\it#1}}}}
  \newcommand{\stanko}[1]{\noindent{\color{Purple} {\bf \fbox{SN} {\it#1}}}}
  \newcommand{\todo}[1]{\textbf{TODO: [#1]}}
  \newcommand{\TODO}[1]{\textbf{TODO: [#1]}}
\fi

\newcommand{\ignore}[1]{}

\usepackage[normalem]{ulem}
\usepackage{multirow}
\usepackage{xspace}

\newcommand{\system}{Storm\xspace}
\newcommand{\systemopt}{Storm-opt\xspace}
\newcommand{\txbench}{TxBench\xspace}
\newcommand{\send}{\texttt{send}\xspace}
\newcommand{\recv}{\texttt{recv}\xspace}
\newcommand{\rdmaread}{\texttt{read}\xspace}
\newcommand{\rdmareads}{\texttt{read}s\xspace}
\newcommand{\rdmawrite}{\texttt{write}\xspace}

\newcommand{\mus}{~\mbox{$\mu s$}}

\usepackage{comment}
\usepackage{cite}

\usepackage{siunitx}
\usepackage{amsmath}
\usepackage{listings}
\usepackage{graphicx}
\usepackage{subfig}
\usepackage{algorithm}
\usepackage[noend]{algpseudocode}	
\usepackage{stackengine}
\usepackage{tabularx}

\usepackage{algpseudocode}
\usepackage{titlesec}
\usepackage{siunitx}
\usepackage[all,defaultlines=3]{nowidow}
\algrenewcommand\algorithmiccomment[2][\normalsize]{{#1\hfill\(\triangleright\) #2}}

\usepackage{array}
\makeatletter
\let\MYcaption\@makecaption
\makeatother

\makeatletter
\let\@makecaption\MYcaption
\makeatother

\usepackage{dblfloatfix}
\usepackage[nolessnomore, italic]{mathastext}
\usepackage[T1]{fontenc}

\usepackage{setspace}
\usepackage{indentfirst}
\usepackage{footmisc}
\usepackage{tabularx}
\usepackage[flushleft,para]{threeparttable}
\usepackage{tabularx,ragged2e}
\usepackage{cprotect}
\usepackage[keeplastbox]{flushend}

\usepackage[us,12hr]{datetime}
\usepackage{booktabs} % For formal tables
\usepackage{mathtools}
\usepackage{xspace}

\newif\ifcameraready
\camerareadytrue
\ifcameraready

\else

\fi

\definecolor{amber}{rgb}{1.0, 0.49, 0.0}
\definecolor{darkgreen}{rgb}{0.0, 0.2, 0.13}
\definecolor{darkbyzantium}{rgb}{0.36, 0.22, 0.33}
\definecolor{darkseagreen}{rgb}{0.56, 0.74, 0.56}
\definecolor{darkspringgreen}{rgb}{0.09, 0.45, 0.27}
\definecolor{dollarbill}{rgb}{0.52, 0.73, 0.4}
\date{}

\begin{document}

\title{\setstretch{0.8}\Large{\textbf{Storm: a fast transactional dataplane for remote data structures}}}


\author{%
{Stanko Novakovic\small{$^1$}}\quad%
{Yizhou Shan\small{$^2$}}\quad%
{Aasheesh Kolli\small{$^1$$^,$$^3$}}\quad%
{Michael Cui\small{$^1$}}\quad%
{Yiying Zhang\small{$^2$}}\\%
{Haggai Eran\small{$^4$}}\quad%
{Liran Liss\small{$^4$}}\quad%
{Michael Wei\small{$^1$}}\quad%
{Dan Tsafrir\small{$^1$$^,$$^5$}}\quad%
{Marcos Aguilera\small{$^1$}}\vspace{2pt}\\%
{\small\it$^1$VMware \qquad $^2$Purdue University \qquad $^3$ The Pennsylvania State University \qquad $^4$ Mellanox \qquad $^5$ Technion}%
\vspace{-5pt}%
}

%\affilVMware

%{Saugata Ghose\affilCMU}\\%
%{Ankit Singla\affilETH}\quad%
%{Pratap Subrahmanyam\affilVMWare}\quad%
%{Onur Mutlu\affilETH$^,$\affilCMU}

\maketitle

% SAUGATA: add page numbers
\ifcameraready
  \pagenumbering{gobble}
\fi
%\else
%  \thispagestyle{plain}
%  \pagestyle{plain}
%\fi
%\newif\ifcameraready
%%\camerareadytrue
%\camerareadyfalse

% Metadata Information
\newcommand{\versionnum}[0]{7.1}

%\fancyhead{}
\ifcameraready
 \thispagestyle{plain}
 \pagestyle{plain}
\else
 \fancyhead[C]{\textcolor{MidnightBlue}{\emph{Version \versionnum~---~\today, \ampmtime}}}%\today, \ampmtime}}}
 \fancypagestyle{firststyle}
 {
   \fancyhead[C]{\textcolor{MidnightBlue}{\emph{Version \versionnum~---~\today, \ampmtime}}}%%\today, \ampmtime}}}
   \fancyfoot[C]{\thepage}
 }
 \thispagestyle{firststyle}
 \pagestyle{firststyle}
\fi

%\setstretch{0.837}
%\renewcommand{\footnotelayout}{\setstretch{0.9}}

%%%%%% -- PAPER CONTENT STARTS-- %%%%%%%%

%\input{title}
\begin{abstract}
RDMA is an exciting technology that
  enables a host to access the memory of a remote host
  without involving the remote CPU.
Prior work shows how to use RDMA to improve the performance
  of distributed in-memory storage systems.
However, RDMA is widely believed to have scalability issues,
  due to the amount of active protocol state that needs to be cached 
  in the limited NIC cache.
These concerns led to several software-based
  proposals to enhance scalability by
  trading off performance.
In this work, we revisit these trade-offs 
  in light of
  newer RDMA hardware and propose new
  guidelines for scaling RDMA.
We show that using one-sided remote memory primitives leads to higher performance compared to send/receive and kernel-based systems in rack-scale environments.
Based on these insights, we design and
  implement \system, a transactional dataplane using one-sided read and write-based RPC primitives.
We show that \system outperforms eRPC, FaRM, and LITE by 3.3x, 3.6x, and 17.1x, respectively, on an Infiniband EDR cluster with Mellanox ConnectX-4 NICs.
\end{abstract}
\section{Introduction}
\label{sec:intro}

RDMA is coming to data centers~\cite{guo:rdma,mprdma-nsdi18,timely:radhika,zhu:sigcomm15,EFA}.
While RDMA was previously limited to high-performance computing environments
  with specialized Infiniband networks, RDMA is now available in
  cheap Ethernet networks using technologies such as RoCE~\cite{rocev1,rocev2} or iWARP~\cite{iwarp}.
The main novelty of RDMA
  is {\em one-sided operations}, which permit an application
  to directly read and write the memory of a remote
  host without the involvement of the remote CPU.
In theory, one-sided operations are supposed to lower latency, 
  improve throughput,
  and reduce CPU consumption.
However, prior work shows that one-sided operations suffer from
  scalability issues: with more than a few hosts, overheads in RDMA can overwhelm the benefits that it provides~\cite{dragojevic:farm,kalia:fasst}.

%In this paper, we examine these issues in light of new hardware
 % developments.
Our first contribution is a study of multiple generations of RDMA NICs
  to understand how hardware evolution addresses (or not) its scalability concerns.
The conventional wisdom is that one-sided RDMA
  performs poorly because of three issues (\S\ref{sec:motiv}).
First, it requires the use of reliable connections, which
      can exhaust the memory cache of the NIC.
Second, one-sided RDMA typically demands virtual-to-physical address translation and
    memory-region protection metadata, which can also exhaust the NIC cache.
Third, one-sided RDMA can incur many network round trips when an application
  wants to chase pointers remotely in dynamic data structures.
  
  %such as linked lists,
  %skip lists, and trees.
%As a result, prior work proposes either reducing the number of
%  reliable connections by sharing it across threads~\cite{farm}, or
%  avoiding reliable connections altogether~\cite{
%
%(inside paper)
%-> kernel solutions (LITE)
%-> QP sharing (FaRM)
%-> avoid connections
%
%We find that the first issue is no longer a concern in rack-scale deployments:

In this paper, we reexamine these problems in light of new and
  better hardware relative to prior work~\cite{dragojevic:farm,kalia:fasst,lite:tsai}.
  We find that some of the problems are mitigated; they are no longer a concern for rack-scale systems of up to 64 machines. 
  Through experiments, we demonstrate that newer hardware efficiently supports
  a significantly larger number of connections than before, eschewing
  the scalability problem for rack-scale.
Furthermore, we argue that connections actually {\em help} performance,
  as they permit delegating congestion control to the hardware and enable one-sided operations.
  % do we show this with some experiment? If so, include a reference
  % to the proper section here.
Thus, systems should use reliable connections as the only
  transport for RDMA communication (\S\ref{sec:principles}).
This is in stark contrast to some previous proposals, such as
  HERD~\cite{herd-rpc-atc16}, FaSST~\cite{kalia:fasst}, and eRPC~\cite{erpc}, which call for
  abandoning reliable connections with one-sided operations
  in favor of the unreliable transport with %traditional
  send/receive operations.

The second issue (virtual address translation and protection metadata) is
  mitigated in newer hardware, but remains.
While future hardware might solve this problem altogether (with
  larger NIC memories and better mechanisms to manage its cache),
  we must still address it today.
Prior solutions are to use huge pages to reduce region metadata~\cite{dragojevic:farm}
  or to access RDMA using physical addresses through a kernel
  interface~\cite{lite:tsai}.
These approaches are effective but have some drawbacks:
  huge pages are prone to fragmentation, while a kernel interface suffers
  from syscall overheads and lock contention issues. In addition, prior research has pointed out that many systems do not aim to allocate memory contiguously, requiring more memory protection metadata~\cite{lite:tsai}.
In this work, we propose enforcing contiguous memory allocation and leveraging the support for physical segments in user-space 
  (\S\ref{sec:principles}): 
We find this approach to greatly reduce region metadata without the concerns of fragmentation or kernel overheads.

%to access
%  contiguous physical addresses from %user-space using a contiguous
%  memory allocator and the hardware support %for physical segments %(\S\ref{somelatersection}).
  
Third issue (round trips to chase pointers) is fundamental, but
  arises only in certain workloads and data structures that require
  pointer chasing.
Prior solutions fall in two categories:
  (1) replace one-sided operations with RPCs~\cite{kalia:fasst,erpc}, so that
  the RPC handler at the remote host can chase the pointers and send
  a reply in a single round trip, 
  or (2) use data inlining and perform larger one-sided reads~\cite{dragojevic:farm}.
In this work, we adopt a new approach
  that performs better than prior solutions:
  the system dynamically determines whether
  to use one-sided operations or RPCs, depending on whether
  pointers need to be chased, and then
  uses the best mechanism. We refer to this hybrid scheme combining one-sided reads and write-based RPCs as \textit{one-two-sided} operations (\S\ref{sec:principles}).
When using RPCs, we employ
  one-sided write operations to transmit the
  RPC request and replies.
%  This approach outperforms two-sided send/receive on modern hardware.

Finally, RDMA is difficult to use as it requires expert knowledge of the low-level protocols and APIs. In this work, we propose a simple, well-understood transactional interface to RDMA. In addition, we propose additional data structure interface allowing developers to specify any remote or distributed data structure in a uniform way. The data structure interface separates the data plane from the data structure itself.
%that we call
%  {\em hybrid one-two-sided
% operations}
  
%We also find that other problems remain even for small deployments,
%  and we propose better solutions to them in the form of a new
Based on our insights, we design and implement a high-speed, transactional RDMA dataplane called \system (\S\ref{sec:design}).
\system can effectively use one-sided operations
  in a rack-scale system, despite prior concerns that they suffer from poor performance~\cite{kalia:fasst,erpc}.
We evaluate \system and compare it against three state-of-the-art
  RDMA systems: FaSST/eRPC~\cite{kalia:fasst,erpc}, FaRM~\cite{dragojevic:farm}, and LITE~\cite{lite:tsai}.

  eRPC is designed to avoid one-sided operations altogether.
We show that \system outperforms eRPC up to 3.3x by effectively using one-sided operations for direct reads and RPCs. Unlike two-sided reads, one-sided reads enable full-duplex \textit{input-output operations per second}(IOPS) rates; no CPU-NIC interaction for processing replies.  
FaRM is designed and evaluated under an older generation of hardware and
  includes a locking mechanism to share connections.
Our measurements show that this mechanism
  is no longer needed
  and produces overhead with newer hardware; 
  we thus improve FaRM by
  removing this locking mechanism and our comparison refers to this improved design.
Our evaluation shows that \system outperforms the improved FaRM up to 3.6x.
Our better performance comes primarily from avoiding large reads in FaRM and 
  instead using fine-grained reads combined with our hybrid \textit{one-two-sided}
  operations. For smaller key-value pairs, FaRM performs significantly better compared to our measurements which are based on 128-byte data items.
Finally, LITE is designed to work in the kernel; we improved LITE by extending it with 
  support for asynchronous operations; our comparison refers to this improved scheme.
Our evaluation shows that \system outperforms the improved LITE up to 17.1x.
Our better performance comes primarily from using user-space operations and a design that is free of dependencies, while we find that
  LITE is bottlenecked by the kernel overheads and sharing among the kernel and user-level threads (\S\ref{sec:eval}).

To summarize, we make the following contributions:

\begin{itemize}

\item We revisit the problems facing one-sided RDMA operations
   in light of new hardware. We perform a detailed experimental study across
   three generations of hardware to understand how its evolution
   addresses (or not) each problem.

\item We build a fast RDMA dataplane called \system, which incorporates the
  lessons we learned from our experimental study. \system provides a transactional API for manipulating remote data structures 
and allows the developer to implement any such data structure using a callback mechanism. \system takes full advantage of one-sided remote primitives using a connected transport and avoids
  lock-based connection sharing that is no longer needed. In addition, \system introduces two new mechanisms,
  user-space contiguous memory allocator and hybrid \textit{one-two-sided}
  operations, to address the problems that remain.

\item We evaluate \system and compare it against eRPC, and improved versions of FaRM and
  LITE, dubbed Lock-free\_FaRM and Async\_LITE. We show that \system performs well in a rack-scale setting
  with up to 64 servers and outperforms eRPC,
  Lock-free\_FARM, and Async\_LITE by 3.3x, 3.6x, and 17.1x in terms of throughput. \system does not trade latency for throughput and provides competitive round-trip times compared to previous systems.

\end{itemize}

Ultimately, \system refutes a widely held belief that one-sided
  operations---the main novelty of RDMA---are inefficient due to
  its scalability issues.

\section{Background}
\label{sec:bg}

\subsection{Remote Direct Memory Access (RDMA)}

\begin{table*}[h!] 

{\footnotesize
\begin{center}
\begin{tabular}{|m{2.3cm} | m{5.7cm}| m{6.2cm} |}
\hline 
\textbf{State:} & \textbf{Includes:} & \textbf{Amount of cached state depends on:} \\
%\hline
%Operating system & \todo{XX} & \todo{XX}\\
\hline 
QP connections & QP metadata, congestion control state & Server count and thread count\\ 
\hline
WQE & Information about a requested operation & Number of outstanding operations\\
\hline
MTT & Virtual-to-physical address translations & Amount of registered memory and page size\\
\hline 
MPT & Buffer ranges and protection keys & Number of registered RDMA buffers\\ 
\hline
\end{tabular}
\end{center}
}
\caption{Sources of transport-level state in RC transport}
\label{table:state}
\end{table*}

RDMA allows applications to directly access memories of remote hosts, with user-level
  and zero-copy operations for efficiency.
Moreover, RDMA offloads the network 
  stack to the Network Interface Card (NIC),
  reducing CPU consumption.
RDMA was originally designed for specialized Infiniband (IB) networks used in high-performance
  computing~\cite{pfister2001introduction}.
%specialized transport layer and a reliable link layer.  
More recently, the IB transport has been   
  adapted for Ethernet networks,
  bringing RDMA to commodity datacenter
  networks~\cite{guo:rdma,zhu:sigcomm15}. Infiniband networks have traditionally been more efficient than RoCE, but now the gap is closing~\cite{personal}.
The IB transport implements an RDMA state 
  machine, handles congestion, and exposes an API to applications.
  
\paragraph{Memory management.} To use RDMA,
applications register memory regions with the NIC, making them available for remote access. 
During registration, the NIC driver pins the memory pages and stores their
  virtual-to-physical address translations in \emph{Memory Translation Tables} (MTTs).
The NIC driver also records the memory region permissions in 
  \emph{Memory Protection Tables} (MPTs).
When serving remote memory requests, the NIC uses MTTs and MPTs to locate the
  pages and check permissions.
The MTTs and MPTs reside in system memory,
  but the NIC caches them in SRAM.
If the MTTs and MPTs overflow the
  cache, they are accessed from main memory via DMA/PCIe, which incurs
  overhead.

%Once the memory registration is %complete, the NIC driver returns a %``region handle'' %(\aasheesh{correct?}), that can be %used by remote clients to access %the registered memory.

\paragraph{Queue pairs.} Applications issue RDMA requests via the IB transport API, known as IB verbs.
IB verbs use memory-mapped control structures called \emph{Queue Pairs} (QPs).
Each QP consists of a Send Queue (SQ) and a Receive Queue (RQ).
Applications initiate RDMA operations by placing Work Queue Entries (WQEs) in the SQ; when operations complete, applications
  are notified through the Completion Queue (CQ).
This asynchronous model allows applications to pipeline requests and do other work while operations complete.

RDMA supports two modes of communication:
one-sided performs data transfers without the remote CPU; two-sided is the traditional send-receive paradigm,
  which requires the remote CPU to handle the requests.
One-sided operations (\rdmaread/\rdmawrite) deliver higher throughput (i.e., IOPS), while two-sided operations (\send/\recv) offer 
  more flexibility as they involve the remote CPU.
%For example, to traverse a linked list in remote memory, multiple {\rdmaread}s are required, while two {\send}s suffice (one to
%  transmit the request, another to transmit the response).
%Based on the application's memory access patterns, one of the two modes of communication is chosen. 
%This paper demonstrates that remote writes can be used to implement a high-throughput two-sided mechanism (i.e., RPC). 

\paragraph{Transports.} RDMA supports different transports;
  we focus on two: Reliably Connected (RC) and Unreliable Datagram (UD).
The RC transport requires endpoints to be connected and the connection to be associated with a QP.
Applications must create one connection (and thus one QP) for each \emph{pair} of communicating endpoints;
  thus, the number of connections (and thus QPs) grows quickly with the cluster size.
For each QP, the system must keep significant state: QP metadata, congestion control state~\cite{timely:radhika,zhu:sigcomm15},
  in addition to WQEs, MTTs, and MPTs
  (Table~\ref{table:state}).
QP state amounts to ${\approx}375B$ per connection~\cite{erpc}.  
UD does not require connections; a single QP allows an endpoint to communicate with any target host.
Thus, UD requires significantly fewer QPs, which saves transport state.
But UD has some drawbacks: it is unreliable (requests can be lost), it does not support one-sided operations, and it
  requires receive buffers to be registered with the NIC, which impacts scalability as we show later.

\subsection{Distributed in-memory systems using RDMA}
Prior work shows how to build
  distributed in-memory storage systems using RDMA~\cite{namdb-vldb16,dragojevic:farm,kalia:fasst,pilaf-atc13,namdb-vldb17,crail,cell-atc16}.
Such storage systems tend to have (i) high communication fan-out; (ii) small data item size, 
and (iii) 
  moderate computational overheads. 
Systems with these properties benefit from RDMA's low-latency and high IOPS rates.
For example, in a transactional store, clients issue transactions with many read/writes on different objects, where
  data is partitioned across the servers~\cite{atikoglu:workload}.
Using RDMA, clients can read/write data using {\rdmaread}s and {\rdmawrite}s or implement lightweight RPCs for that purpose, reducing
  the end-to-end latency and improving throughput.

\section{Motivation}
\label{sec:motiv}

\subsection{Problem statement}
Our main goal is to use RDMA efficiently and scalably in a rack-scale setting.
While some companies have mega-deployments with thousands of machines, the vast bulk of 
enterprises use rack-scale deployments, consisting of one or a few racks
  with up to 64 machines in total; that is our target environment.
Prior work has shown that RDMA-based distributed storage systems do not scale well in
  these settings~\cite{dragojevic:farm,erpc,lite:tsai}.
As we add more machines and increase their memory, the amount of RDMA state increases.
%RDMA offloads significant network stack functionality to the NIC.
For good performance, the active RDMA state must be in the NIC's SRAM cache,
  but this cache is small and can be exhausted with a few remote peers~\cite{dragojevic:farm}.
%even for a handful of nodes become a key performance obstacle.
When that happens, RDMA state spills to CPU caches and main memory, requiring expensive
  DMA operations over PCIe to access it.
%These DMA operations are
%especially problematic for throughput-bound datacenter applications, which is the focus of this work.
PCIe latency adds 300-400ns on unloaded systems to several microseconds on loaded systems~\cite{personal,pcie-sigcomm18}.
%(similar to DRAM)~\cite{goldhammer2008understanding}.
%, whereas on-chip SRAM latency adds less than 10ns.
These DMA overheads are exacerbated with
  transaction processing workloads, which have high fan-out, fine-grained accesses.
%Table~\ref{table:state} shows the transport-level state  in RDMA, including: (i) QPs storing connection state, requests, and completions; (ii) MTTs storing virtual-to-physical address 
%translation information; and (iii) MPTs keeping track of registered memory regions and
%protection keys. 

To mitigate this problem, several software solutions have been proposed. 
We discuss them next and argue that recent RDMA hardware  changes the design trade-offs originally
  envisioned, particularly in distributed in-memory storage systems.

\begin{comment}
The goal of this work is to develop high throughput and scalable transaction processing systems using RDMA.
Prior work has shown that RDMA has scalability issues and have proposed various techniques to mitigate these issues.

\end{comment}

\subsection{Shortcomings of prior art}

\begin{comment}
\begin{itemize}
    \item LITE onloads address translation and protection. Introduces system calls in critical path. 
    \item FARM uses 2GB pages and onloads QP sharing. QP contention limits scalability.
    \item eRPC uses RPCs over UD. Requires posting Recv buffers and onloaded congestion control, limiting throughput and scalability. 
\end{itemize}
\end{comment}

\begin{comment}
\begin{table}[t]

\setlength{\tabcolsep}{2pt}
\centering\begin{tabular}{|c|c|c|c|c|c|}
	\hline
	system 	& 	bare-metal 	& 	reliable 	& 	scalable 	& 	target \\
	\hline\hline
	FaSST/eRPC		&  no  & yes & yes & QP\\
	%\hline
	\cline{1-1} \cline{2-2} \cline{3-3} \cline{4-4} \cline{5-5}
	FaRM 	&	no & yes & no & QP, MTT \\
	\cline{1-1} \cline{2-2} \cline{3-3} \cline{4-4} \cline{5-5}
%	(projected) &		&		&		& \\
	LITE		&  no  & yes & yes & QP,MTT,MPT\\	
	\hline	
\end{tabular}
\caption{Systems aiming to improve RDMA scalability.}
\label{tbl:related}
\end{table}
\end{comment}

%In Table~\ref{tbl:related}, we

We focus on three systems trying to address RDMA scalability issues in software. 
These systems, we argue, do not fully leverage more recent RDMA hardware.

%In addition,
%we discuss the performance impact of shared memory abstractions.

%Finally, LITE currently does not support asynchronous RDMA, which is necessary for achieving
%igh throughput. Asynchronous RDMA can be implemented, but will introduce additional kernel 
%traps on the data path. 

\paragraph{Systems using one-sided operations.}
FaRM~\cite{farm2:aleks} and LITE~\cite{lite:tsai} use one-sided operations
  and try to reduce the number of QPs by sharing them across groups of threads.
%This has been 
%necessary as RDMA has traditionally been prone 
%to thrashing the QP state as the number of 
%connections grows.
%Without sharing, QP state grows with the number of connections (about 375B~\cite{erpc} per connection) and may thrash the NIC cache.
%The QP state amounts to 375B~\cite{erpc} per connection and for best performance all of the state needs to be cached on the NIC. As this is not feasible even at the 
%scale of a few tens of nodes, one solution is 
To share, these systems use locks,
%QP sharing through locking reduces QP state for a given number of connections and delivers better NIC cache performance.
%to have fewer QP connections and share them 
%through locking.
%However, this has been proven
  but locking degrades throughput~\cite{dragojevic:farm}.
Also, FaRM uses large reads to reduce the number of 
  round-trips when performing lookups, limiting maximum throughput.
%We find this to be a 
%limitation for high lookup rates.
%To minimize
%the number of round-trips, FaRM applies the hopscotch technique to store a set of 

\paragraph{Unreliable datagram transport.}
%There are several transactional systems taking advantage of RDMA to achieve high throughput. 
%FaRM~\cite{dragojevic:farm} relies on one-sided \rdmareads to perform remote lookups.
%\rdmareads require a connected transport, which requires QP sharing to minimize NIC cache thrashing. 
%While FaRM~\cite{dragojevic:farm} uses locking to share QPs to reduce NIC cache thrashing, 
Another way to reduce QP state is to use the UD transport, as in FaSST/eRPC~\cite{erpc,kalia:fasst}. With UD,
  a thread uses just one QP to talk to all the machines in the cluster.
%, that way minimizing the amount of QP state.
However, UD precludes the efficient one-sided operations (\rdmaread/\rdmawrite), requires 
  application-level retransmission, and requires application-level
  congestion control, all of which limit maximum throughput (\S\ref{sec:eval}).
Furthermore, we show that managing receive queues in UD impacts scalability.
%because of the 
%send/recv model where receive buffer have to be
%registered (i.e., posted) with the NIC, eRPC does not provide perfect scalability.

\paragraph{Kernel-space RDMA stacks.}
LITE~\cite{lite:tsai} provides a kernel interface for RPCs and
  remote memory mapping.
Although LITE eliminates MTT/MPT overhead in the NIC, it adds additional overhead due to frequent system calls which are now somewhat more costly due to recent kernel patches (i.e., KPTI, retpoline)~\cite{meltdown}.
Moreover, LITE operations
  are blocking, which limits concurrency and throughput.
We extend LITE with asynchronous \rdmareads and RPCs to improve its throughput. 
This version achieves 2$\times$ higher throughput for a single thread (\S\ref{sec:eval}), 
  but the maximum IOPS with multiple threads remains small compared to RDMA on a modern NIC.
We find this occurs because of serialization and lock contention in LITE.

\begin{figure}[t]
    \includegraphics[width=0.47\textwidth]{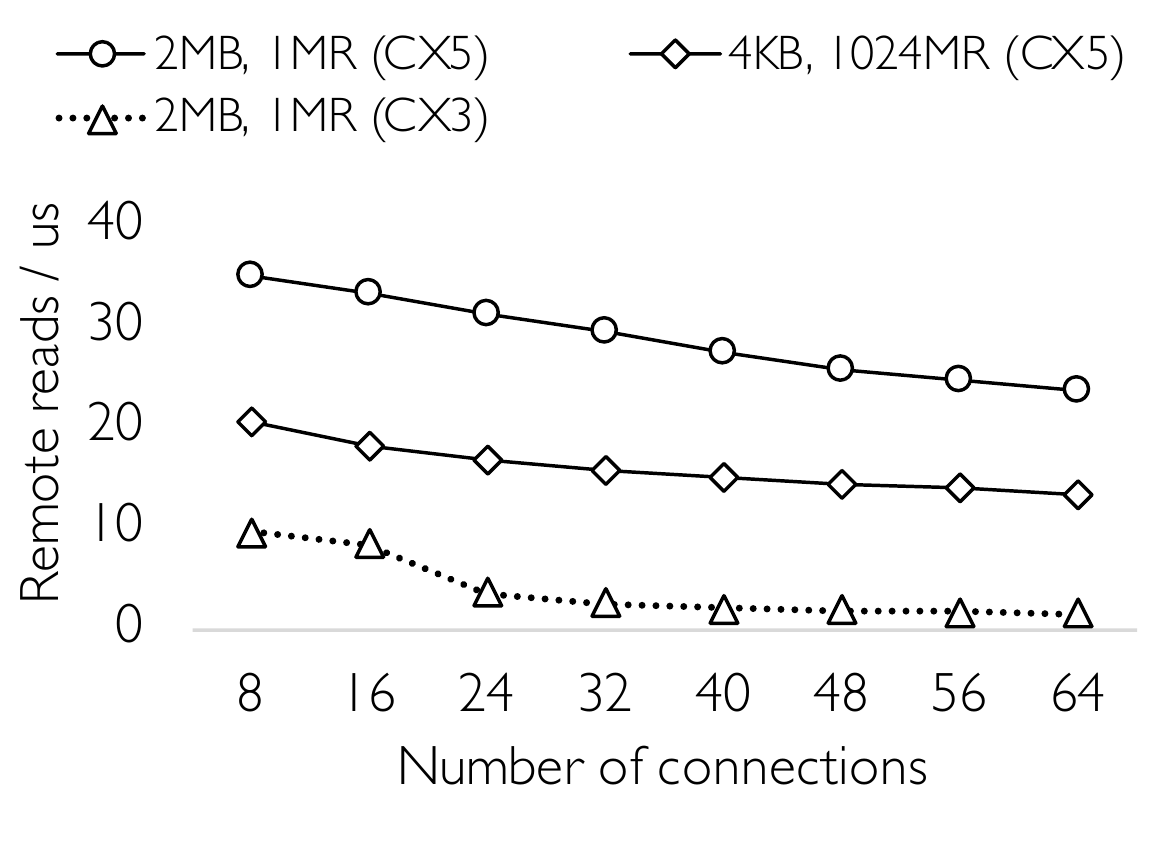}
    \caption{CX3 vs. CX5 comparison. CX5 provides better performance even when using 4KB pages (as opposed to 2MB) and registering a larger number of memory regions (1024MR)}
    \label{fig:motivation}
\end{figure}

%To minimize the amount of QP metadata, both FaRM and LITE share QPs across a group of threads.
%However, sharing QPs leads to lock contention and degraded throughput. 

%is another example of a distributed object store, which relies on RPC rather 
%than remote reads and writes. A key design decision i
%FaRM uses a variant of the 
%two-phase commit protocol to provide fully serializable transactions.

%FaSST avoids registering the 
%entire datastore memory at the cost of additional copying for each RPC response.  

\subsection{Revisiting RDMA hardware capabilities}
\label{sec:revisit}
%Prior art was developed on older versions of RDMA hardware.
%Most recent versions of RDMA hardware deliver significantly better performance.
%Present CX-3 vs CX-4 vs CX-5.
%\begin{itemize}
%    \item Larger caches
%    \item Better prefetching and caching policies
%    \item Larger number / improved processing units
%    \item Higher bandwidth
%    \item Physical segment support
%    \item More efficient transport protocols (i.e., DC)
%\end{itemize}

FaRM, FaSST/eRPC, and LITE were designed for old NICs (CX3) with very limited processing and memory resources. 
Their design choices (e.g., QP sharing and software address translation) improve performance on such NICs
  but underutilize the capabilities of newer hardware (CX4 and CX5).
Figure~\ref{fig:motivation} tries to capture a significant performance gap between CX3 and CX5 Mellanox RoCE NICs.
It shows the throughput per machine for a workload that performs random 64-byte remote \rdmareads on 20GB of memory (2MB page sizes).
For CX5, we also show the performance when MTTs and MPTs are larger; \textit{4KB,1024MR (CX5)} uses 4KB pages and breaks the 20GB of buffer space into 1024 smaller RDMA memory regions (MR). On the X-axis we vary the number of established connections between the source and destination servers. The RoCE hardware is described in Table~\ref{table:evalplatforms}.

We draw three conclusions from Figure~\ref{fig:motivation}: (i) CX5 significantly outperform CX3; (ii) CX5 scales better with the number of connections than CX3. We measure the throughput reductions going from 8 to 64 connections to be: 83\%, 42\%, and 32\% for CX3, CX4, CX5, respectively; (iii) MTT and MPT remain a significant overhead with many memory regions and large page counts.  
Finally, we find that CX5 throughput becomes constant at around 10000 QP connections after reaching zero cache hit rate. The constant throughput that we measure is around 10 $reqs/\mus$, which is equal to the maximum throughput a CX3 can provide (when there is no contention).
Next, we list a number of factors that drive the better performance of modern NICs.
%We find that CX4/CX5 NICs has high read throughput even with thousands of QP connections, which amounts to hundreds of machines.
%The superior performance of CX4/5 arises from larger NIC caches (${\approx}$2MB) and better cache management policies~\cite{erpc}.

\paragraph{Larger cache sizes, better cache management.}
CX4/5 has larger caches (${\approx}$2MB)~\cite{erpc} for RDMA state, reducing the number of PCIe/DMA operations on
  system memory.
%While improving cache sizes is one way to reduce NIC cache misses, it is not a scalable option. 
Moreover, these NICs can better utilize their cache space, with improved
   prefetching, higher concurrency, and better cache management~\cite{personal}. Such optimizations allow a modern RDMA NIC to deliver competitive throughput even when there are virtually no cache hits on the NIC. 
   
%  \yiying{maybe mention here that the increased cache size (and better processors etc.) does not come with a huge increase in monetary cost, or people may ask if these benefits come with extreme high cost in hardware, which is not the case}
%allow caches to be more effectively utilized even when the NIC caches are being overwhelmed.

\paragraph{More and improved processing units.} Modern RDMA NICs are equipped with increasingly powerful Processing Units (PUs).
This allows NICs to issue more requests in parallel, which in turn increases throughput and
  hides PCIe latency to fetch data on cache misses~\cite{personal}.
%
%, providing
%  higher IO parallelism and throughput,
%In RC, each QP connection is mapped to one processing unit, ensuring in-order delivery and avoiding contention 
%among the PUs~\cite{personal}. By using many QP connections and hence many processing units, 
%As a result, the NIC can issue more requests in parallel.
This obviates the need for various aggregation techniques 
  and data layout optimizations used previously. For a sufficient number of active QP connections (each mapped to a single PU), a CX5 RoCE delivers close to 40 million reads per second (no contention).

\paragraph{Physical segment support.} CX4 and CX5 support physical segments with bound checks. Unlike CX4/CX5, CX3 only supports registering the entire physical memory of a server, precluding physical addressing from user-space. This mechanism 
  bypasses virtual-to-physical translation and reduces the MPT and MTT sizes.
This is important for hosts with large persistent-memory systems
  with tens of TBs to a PB of memory~\cite{3dxpoint-news,swift-hotos17}. In such systems, even 1GB pages could lead to large
  MTTs (e.g., 100TB would require close to 1MB of MTT with 1GB pages). Physical segments support arbitrarily large memory regions with just one MPT entry and no MTTs.

\paragraph{Efficient transport protocols.} QPs in RC consume 375B per connection~\cite{erpc}, and RC requires many connections,
  which can overwhelm the NIC caches.
%MKA: we already said the stuff below, this is repetitive
%One option is to not use connections and rely on UD. However, that alternative requires the application to handle packet losses and manage congestion control. In addition, UD precludes using \rdmaread/\rdmawrite and requires managing a receive queue, which we show later is a scalability limitation.
Modern NICs provide a new transport called Dynamically Connected (DC)~\cite{IB-DCT}, which can share a QP connection across multiple hosts, thereby
  reducing the amount of QP state.
%Managing QP sharing in hardware promises to perform and scale better than managing it in software. 
%DC currently suffers from breaking and establishing connections when issuing requests to different machines and is not available for RoCE~\cite{personal}. 
DC is not available for RoCE and suffers from frequent reconnects which diminish its purpose~\cite{kalia:fasst}. In this paper we focus on the RC transport.
As we show, RC scales well on clusters with up to 64 hosts.
%, which we find is the common case for RDMA deployments.

\subsection{Revisiting prior work on improved RDMA}
A key contribution of this paper is to show that on modern NICs, one-sided primitives can outperform alternatives for
moderate cluster sizes (tens of machines), even when the NIC caches are being thrashed. For instance, on CX5(RoCE) it takes on the order of 2500 to 3800 connections for \rdmareads to match the performance of UD-based send/receive~\cite{kalia:fasst,erpc}. We expect the break-even point to increase in the future through the improvements mentioned in \S\ref{sec:revisit}.
We argue that one-sided remote
reads and writes are best for building low-latency, high-throughput, and low CPU utilization systems. 

While FaRM~\cite{dragojevic:farm} also championed the use of one-sided primitives, we differ from them in other ways. For example, rather than sharing QP connections to minimize the amount of QP state, we propose establishing one connection for each ``sibling''
  pair of threads (threads with
the same local ID running on distinct machines) for a total of $2\times m\times t$ connections per machine, where $m$ is the number of machines and $t$ is the number of threads per machine.
We show that this model can scale well on modern hardware. Also, rather than transferring larger quantities of data, the ultra-high IOPS rates of modern RDMA NICs allow for high-throughput fine-grain access. Additionally, we propose leveraging both \rdmareads and \rdmawrite-based RPCs.
The two primitives are efficient for different operations, \rdmaread for simple lookups, RPCs for complex operations such as long linked-list traversals. Finally, we show how physical segments, an RDMA feature championed in LITE~\cite{lite:tsai}, can be used from user-space.

\section{Design principles}
\label{sec:principles}

%\yiying{needs some overview discussion on the principles. e.g., what are these design principles for? what's our design goals? are we desigining it just for transaction system? or can these principles be applicable to building distributed rdma-based systems in general? what's our targeted scale? why do we choose that scale? maybe also mention a bit about challenges (e.g., why is it not enough to just porting existing systems to newer generation of NICs)}

We propose five design principles for
  RDMA-based distributed in-memory systems.
We later show that these principles permit scaling well
  in a rack-scale setting with up to 64 servers.
The principles are as follows:
\\
\paragraph{1. Simple transactional interface for any data structure.} Programming directly on top of RDMA API (i.e., Infiniband verbs) is difficult. Even developing a basic client-server application requires expert knowledge of the low-level Infiniband implementation and the NIC. First, the developer must decide which flavor of the Infiniband transport to use and understand its properties. Then, they have to initialize and register the queue-pairs (QP), as well as the memory, with the NIC. On RDMA, the low-level QPs are used directly by the application to access remote memory. For each access, the developer must provide detailed, low-level information about the requested operation, as this is passed directly to the NIC in form of a work queue entry (WQE). Finally, the verbs API is inherently asynchronous and complex, but the developer must use it in order to make the most out of RDMA. Alternatively, this work proposes a simple transactional API to RDMA, where any remote data structure can be manipulated using transactions. In abstract terms, for each item the user should just specify the following:

\begin{itemize}
	\item Object ID: Identifies an instance of a data structure.
	\item Item ID (key): Identifies a specific item in the data structure (optional depending on the data structure).
	\item Operation type (opcode): Identifies a data structure operation that the user requested
\end{itemize}

In addition, similarly to previous systems~\cite{kalia:fasst}, we propose using a lightweight user-level thread scheduler to take advantage of the asynchronous RDMA API, while providing blocking semantics to the developer. 

\paragraph{2. Leverage RC connections.}
As we mentioned,
   RC has a scalability cost: it consumes more transport-level state than UD, 
   which can lead to NIC cache thrashing.
We show that new hardware---with larger caches, better cache management, and
  more processing units---changes the trade-off in favor of RC in rack-scale deployments.
That is, the cost is more than offset by the many benefits of RC:
  (1) RC allows lightweight one-sided primitives (\rdmaread/\rdmawrite)
      that have lower CPU utilization and achieve higher IOPS;
  (2) RC offloads retransmissions from the CPU to the NIC, and
  (3) RC offloads congestion control as well.
These benefits not only improve performance, but also simplify application
  development (developers need not worry about resending requests
  and rate-limiting transmissions).
%
% While UD reduces the transport-level state as it requires only a single QP per application thread, RC delivers better performance, even at the scale of 10--100 nodes.
% This is because RC allows the use of one-sided primitives (\rdmaread/\rdmawrite), which are
%   lightweight, have lower CPU utilization, and deliver higher IOPS than two-sided primitives (\send/\recv).
% RC offloads retransmissions and congestion control from the CPU to the NIC, further reducing CPU utilization and simplifying application development.
%
We also show that RPCs should be implemented with RC rather than UD, by
  using RDMA \rdmawrite. 
  
  %We show
  %that using \rdmawrites delivers better performance than using \send/\recv.
% While RC requires a larger transport-level state (due to QPs), the downside is mitigated in
%   newer RDMA hardware with larger caches and
%   better cache management.
% Thus, to obtain highest performance,
%   RDMA systems should use RC instead of UD. 

%(i) higher throughput of one-sided primitives and (ii) offloaded congestion control

%It is thus important to reduce this metadata
%  to reduce cache pressure, avoid
%  expensive DMA operations, and thereby %improve throughput and scalability.
  
\paragraph{3. Minimize RDMA region metadata.}
While new hardware addresses NIC cache state concerns for RC, another issue
  remains: cache state for MPTs and MTTs.
Large memories result in excessive
  address translation and protection metadata
  that exhaust the NIC's cache.
This is concerning for applications that allocate memory
  progressively as a large number of smaller chunks
  (e.g., Memcached allocates 64~MB chunks) that have to be registered as separate RDMA regions~\cite{lite:tsai}.

To address this problem, we use two techniques.
First, we minimize the number of registered RDMA memory regions by
  using a contiguous memory allocator~\cite{dragojevic:farm}.
Such an allocator requests large chunks of memory from the kernel and manages small object allocations.
Thus, we only register a small number of large chunks that we expand and shrink dynamically as the application allocates/deallocates memory, minimizing MPTs. The system could then use on demand paging to repurpose unused pages (currently only works for 4KB pages).
 
Second, to reduce the memory translation table metadata (MTTs), we propose
  using {\em physical segments},
  a feature available in newer RDMA NICs such as 
  CX5~\cite{mellanox-phy-seg}.
Physical segments export physical memory with user-defined bounds with little MTT overhead, and
  this feature is available in user-space.\footnote{This
  is different from LITE's approach to export all of physical memory and enforce protection
  in the kernel.}
%This is similar to Direct Segments aiming to reduce translation overheads on the CPU~\cite{direct-seg-isca13}.
Physical segments were intended for   
  single-tenant use; using them in a host
  with many tenants requires care to
  avoid security issues when exposing
  physical memory.
We propose a solution to these issues, by
  mediating the registration of physical
  segments by the kernel.
This approach is secure and imposes minimum
  overhead since kernel calls are off the
  data path.
Moreover, this approach is more efficient than using huge pages (2MB or 1GB)
  to reduce the MTTs~\cite{dragojevic:farm}.
Huge pages lead to fragmentation and
  wastes memory~\cite{ingens-osdi16,Panwar-asplos18}, and may not suffice:
    for large memories with 100s of TBs, even 1GB pages result in large MTTs.
    In addition to the security concerns, physical segments require the use of Linux CMA~\cite{linux-cma}. Thus, it is important to limit the number of physical segments on a machine, as Linux CMA may not be able to efficiently handle multiple growing regions that need to be physically contiguous.

\paragraph{4. First use fine-grain \rdmareads, then switch to RPCs using \rdmawrite.}
One-sided \rdmareads deliver high IOPS for simple lookups~\cite{dragojevic:farm,pilaf-atc13,cell-atc16,namdb-vldb17}.
However, they are less efficient to access
  data structures with cells and pointers,
  such as skip lists, trees, and graphs, which
  require pointer-chasing.
Thus, prior work proposes two alternatives: 
  (1) use RPCs implemented with \send/\recv verbs~\cite{erpc,kalia:fasst} or
  (2) change the data structure to combine many cells and fetch more data at a time~\cite{dragojevic:farm}.

With new hardware, we show that the best approach is as follows.
First, use one-sided \rdmareads to fetch one cell at a time.
Our evaluation shows that combining cells and fetching more data
%  (approach (2) above)
  results in lower throughput.
Second, if the one-sided \rdmaread reveals that we must chase pointers, switch to using RPCs.
We call this hybrid scheme {\em one-two-sided} operations.
Furthermore, we show that RPCs can be implemented efficiently using one-sided {\rdmawrite}s, not \send/\recv.

\paragraph{5. Resize and/or cache.}
For remote reads to be effective, most data structure operations should require one round trip in the common case. Otherwise, RPCs are proven to be more effective~\cite{kalia:fasst}. One round trip 
per operation is hard to achieve, especially with pointer-linked 
data structures. This work proposes a simple approach, which is to trade abundant memory for fewer round trips with one-sided operations.
There are two ways to achieve this trade: (i) clients could cache 
item addresses for future use, as in DrTM+H~\cite{hybrid-osdi18} and/or (ii) for hash tables, when RPC usage becomes excessive, one should resize the data structure (e.g., by adding buckets to a hash table) to keep the occupancy low and reduce pointer chasing due to collisions, while retaining one-cell transfers. We claim that the amount of consumed memory is not significant, especially in the face of high-density persistent memory technologies. For example, in (i), for one billion items, each client machine would have to dedicate 8GB of memory for the whole key space, and not all key-value addresses have to be cached for remote reads to be effective. Clients should be able to perform version checks for retrieved data items to make sure the cached addresses are still valid. The versions require some additional storage on the client. For (ii), we find that keeping the occupancy below 60-70\% is sufficient to emphasize the performance benefits of one-sided reads. Nevertheless, we are looking into ways to repurpose the unused portions of allocated memory. 
%We envision a special type of a memory allocator capable of temporarily repurposing the unused memory.  

\section{Design and Implementation of Storm}
\label{sec:design}
%\subsubsection{\system vs eRPC vs LITE}

Following our principles (\S\ref{sec:principles}), we propose \system, a high-performance RDMA dataplane for remote data structures.
\system is designed to run at maximum IOPS rate of the NIC by using RDMA primitives and by minimizing the active protocol state. \system exposes a transactional API for manipulating remote data structures.

\begin{figure}[t]
    \includegraphics[width=0.47\textwidth]{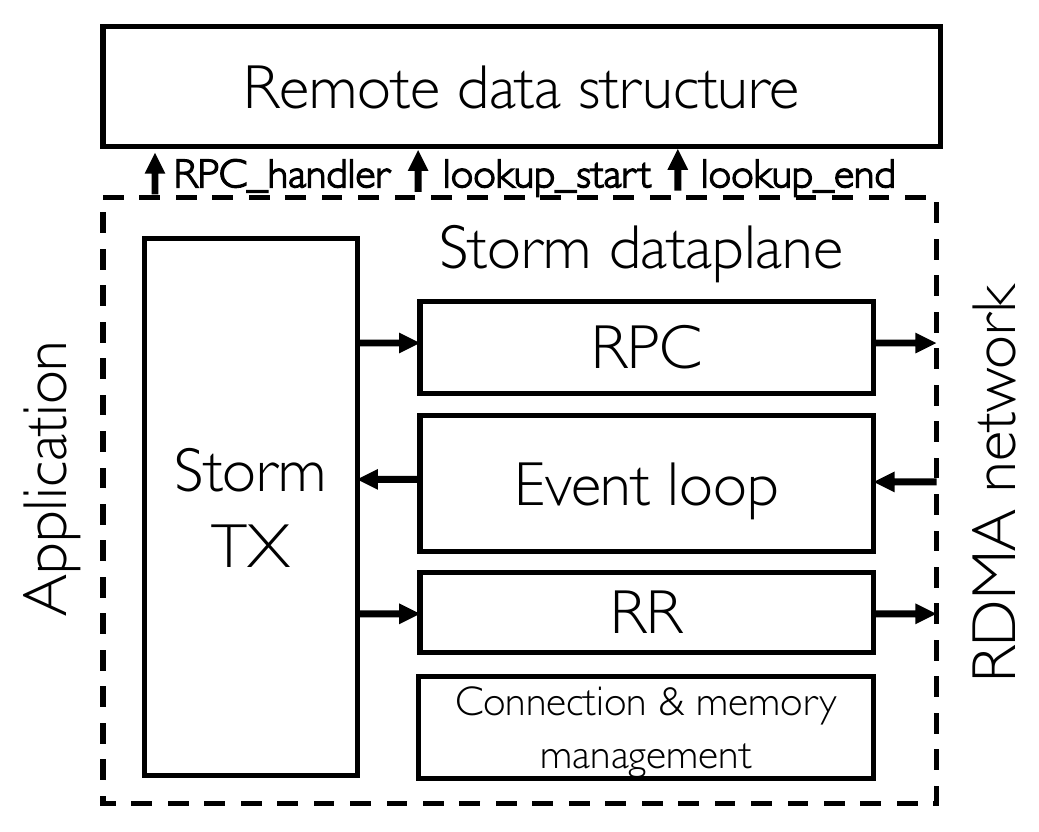}
    \caption{Storm high-level design. Independent pipelines for remote reads and RPCs. A single event loop processing all completions. Data structure completely independent of the data plane. The developer implements the data structure interface consisting of three callback functions.}
    \label{fig:storm_design}
\end{figure}

Figure~\ref{fig:storm_design} shows the high-level design of Storm. Two independent 
data paths, one for RPCs and one for one-sided reads (RR), process remote requests 
coming from the local process.
The event loop processes inbound requests coming from remote processes and all event 
completions. The Storm TX module provides a transactional API 
to the user by leveraging the data structure API and the RPC/RR data paths 
to execute distributed transactions. In this section, we discuss the following: (i) memory allocation, 
(ii) RPC implementation based on remote writes, (iii) Storm TX API and the data structure API, (iv) the Storm transactional protocol, (v) hash table as an example remote data structure, and (vi) concurrency.
%We present its salient features next.
%While the use of one-sided primitives leads to NIC cache thrashing in certain scenarios, we observe that \system outperforms prior designs by up to an order of magnitude 
%Still, the use of one-sided primitives in Storm often leads to thrashing of the protocol state from the NIC cache. However, we find that even on 64-node deployments, Storm outperforms the previous designs by up to an order of magnitude. 

%Storm minimizes the amount of active protocol state and leverages %both one-sided reads and write-based RPCs to achieve maximum %throughput.   

%The result is that all of the experiments presented in this paper %are IOPS bound, except for TATP, which is more complex in nature %(i.e., runs transactions as opposed to simple lookups).

%The first step \yiying{not sure if this is the first...} in designing a high-throughput and scalable storage system on RDMA is efficient memory management that minimizes the amount of translation and protection state on the NIC.

\subsection{Contiguous memory regions}
To achieve best performance, we must manage memory efficiently in RDMA.
%Managing memory efficiently in RDMA is necessary for best performance.
Earlier, Figure~\ref{fig:motivation} showed that large MTTs and MPTs leads to significant performance degradation, even on modern hardware. Thus, Storm aims to allocate 
virtually contiguous memory when possible to minimize the number of registered RDMA regions. By doing so, Storm minimizes the MPT state. In addition, Storm can allocate physically contiguous memory and expose it as one physical segment~\cite{mellanox-phy-seg}, requiring only a single MTT and one MPT entry. The physical segment support for RDMA 
cannot be used in untrusted environments, as any user with the right capabilities can register any part of the local physical memory and access it through RDMA (e.g., access to local kernel memory through a loopback). In
Storm, we require from all applications to register physical segments through the OS kernel.
This registration is not a performance bottleneck as it is not on the critical path. With sufficiently large pages, physical segments may not be necessary. Thus, in most of our experiments we do not use them. However, future storage-class memory systems with PB of memory will require support for physical segments.
%%We show that 
%physical segments reduce address translation state to a minimum (\S\ref{sec:eval}) , which %is important for future storage-class memory systems with TBs to PB of memory. 
%Figure~\ref{fig:polling} shows how Storm allocates a single memory region for communication %purposes for the entire machine. 

%Moreover, we aim to manage memory (completely?) in the user space to avoid kernel trap costs.
%The buffer is divided into many small buffers, one for each adjacent server, thread, and concurrent request.

%In Storm, all the memory is contiguous and application memory is no exception. Within each machine, Storm maintains one data buffer and exports it into the global address space using RDMA. Such contiguous memory allocation minimizes the MPT state. 

%\yiying{hmm... you need more explanation here about the ``new security model''. what's the se}
%The physical segment feature is intended for expert, single-tenant %use and hence is done from user space by the application code. 

%For reads, the overhead of polling is constant, as only k buffers need to be polled, where k is the number of concurrent requests.

\subsection{Remote write-based RPCs}
\system leverages the \textit{rdma\_write\_with\_imm} (RDMA write with immediate) primitive to send and receive messages. This primitive allows the client to prepend a custom header to each message, which is useful for communicating additional information about the sender (e.g., process ID, coroutine ID, etc). More importantly, \textit{rdma\_write\_with\_imm} enables scalable polling on the receiver; the receiver posts onto the receive queue (just like when using send/recv) and receives a notification via a receive completion queue for each received message. This way the receiver does not have to poll on the message buffers. In addition, to avoid polling on multiple receive completion queues, the IB verbs interface permits using a single receive completion queue, independent of the number of senders. We use this messaging primitive to 
implement fast and scalable RPCs in Storm.

\begin{algorithm}
\caption{Processing a read-set item}\label{alg1}
\begin{algorithmic}[1]
\\\textbf{Input:} Data structure object ID, key, size
\\\textbf{Output:} Data item from remote memory
%\State determine workload read to write ratio                                              

\State $success$ $\leftarrow$ $false$
\State $region\_id,offset$ $\leftarrow$ $lookup\_start(object\_id, key)$ 
%\vspace{3mm}
\If{$region\_id$ $\not=$  $-1$}
\State $buffer$ $\leftarrow$ $remote\_read(region\_id, offset, size)$
\State $success$ $\leftarrow$ $lookup\_end(buffer, key)$
\EndIf

\If{$success$ $\not=$ $true$}
\State $buffer$ $\leftarrow$ $rpc\_send(key, READ)$
\State $success$ $\leftarrow$ $lookup\_end(buffer, key)$
\EndIf

\end{algorithmic}
\end{algorithm}

\subsection{\system API}
At a high level, \system exposes an intuitive and well-understood transactional API for manipulating remote data structures (Table~\ref{table:api}). A client can simply add to 
read set/write set and commit transactions once it's finished. Storm's event loop is 
invoked periodically to process event completions. 

Internally, Storm provides the following programming model for remote data structures:
The developer implements three functions and registers them as callbacks with the Storm
dataplane. The functions are listed in Table~\ref{table:apicb}. These functions are implemented as part of the remote data structure. \textit{rpc\_handler} is used for lookups on the owner (receiver) side. Locks and commits are also implemented in this handler.
\textit{lookup\_start} is the remote lookup handler for looking up a remote data structure's metadata on the client side. This metadata could be cached data structure addresses or simply a guess for an object's address based on hash. Algorithm~\ref{alg1} shows at a high level how the Storm dataplane on the client side processes each request from the read-set. It first invokes 
\textit{lookup\_start} to get the RDMA region ID and offset where the requested item may reside. If the region ID is a positive value, the client uses the returned information to look up the item using a remote read.

When a lookup is finished, the client invokes \textit{lookup\_end} to check the validity of the 
returned data. If the data is not valid: for example, the read key does not match the requested key, the client issues an RPC. \textit{lookup\_end} may decide to cache the address of the returned object for future use. This depends on the remote data structure implementation. With RDMA, we cannot afford additional remote reads, as this will hurt performance. Future faster interconnects may change this trade-off. Invoking \textit{lookup\_end} is necessary for lookups using remote reads, but it is also invoked after every RPC lookup, so that the data structure can store the returned address for future use. \textit{lookup\_end} may return false even after the RPC call if, for example, the item does not exist.

\begin{comment}
\system exposes an intuitive API that is common for datacenter applications.
Table~\ref{table:api} shows the core functions. \system exposes one primitive for data transfers (\texttt{storm\_req\_data}) that internally uses both remote reads and write-based RPCs. The developer may provide hints as to 
which primitive should be used. The RPC interface is also exposed to the developer (\texttt{storm\_send}) for explicit messaging purposes. 
To achieve maximum concurrency, one can manually schedule remote operations by invoking \system's asynchronous API and periodically checking for completions (\texttt{storm\_eventloop}).
FaRM and eRPC leverage callback continuations, which 
is a complex programming model to use, but incurs minimal overhead.
In our workloads, we use user-level threads (a.k.a. coroutines~\cite{boost-coroutine}), which are simpler to use and incur less than 20 ns per context switch~\cite{kalia:fasst,hybrid-osdi18}. 
%In our workloads, we use coroutines~\cite{kalia:fasst,boost-coroutine}), which are simpler to use and incur less than 20 ns per coroutine context switch~\cite{kalia:fasst}.
\system API is fully lock-free, so multi-threaded programs do not suffer from lock contention.
\end{comment}

\begin{table}[t!] 
\footnotesize
\centering 
\caption{Storm API}
\label{table:api}
\begin{tabular}{m{3.2cm}  m{4.3cm}}
%\hline 
\textbf{API} & \textbf{Description} \\
%\hline
%Operating system & \todo{XX} & \todo{XX}\\
\hline 
%  \\ %\hline
%\cline{1-1}\cline{2-2}\cline{3-3}
storm\_eventloop & process requests and completions\\
%\cline{1-1}\cline{2-2}\cline{3-3}
storm\_start\_tx & start a new transaction \\
%\cline{1-1}\cline{2-2}\cline{3-3}
storm\_add\_to\_read\_set & add an item to read set\\
%\cline{1-1}\cline{2-2}\cline{3-3}
storm\_add\_to\_write\_set & add an item to write set\\
%\cline{1-1}\cline{2-2}\cline{3-3}
storm\_tx\_commit & commit a transaction\\
%\cline{1-1}\cline{2-2}\cline{3-3}
storm\_register\_handler  & register a callback handler (Table 3)\\
\hline
\end{tabular}

\end{table}

\begin{table}[t!] 
\footnotesize
\centering 
\caption{Data structure API}
\label{table:apicb}
\begin{tabular}{m{3.2cm}  m{4.3cm}}
%\hline 
\textbf{API} & \textbf{Description} \\
%\hline
%Operating system & \todo{XX} & \todo{XX}\\
\hline 
%  \\ %\hline
rpc\_handler & local RPC handler\\
lookup\_start & get data item region ID and offset\\
lookup\_end & check if successful and cache\\
\hline
\end{tabular}

\end{table}

\subsection{\system transactional protocol}
\begin{figure}[t] %h!
    \includegraphics[width=0.47\textwidth,height=5cm]{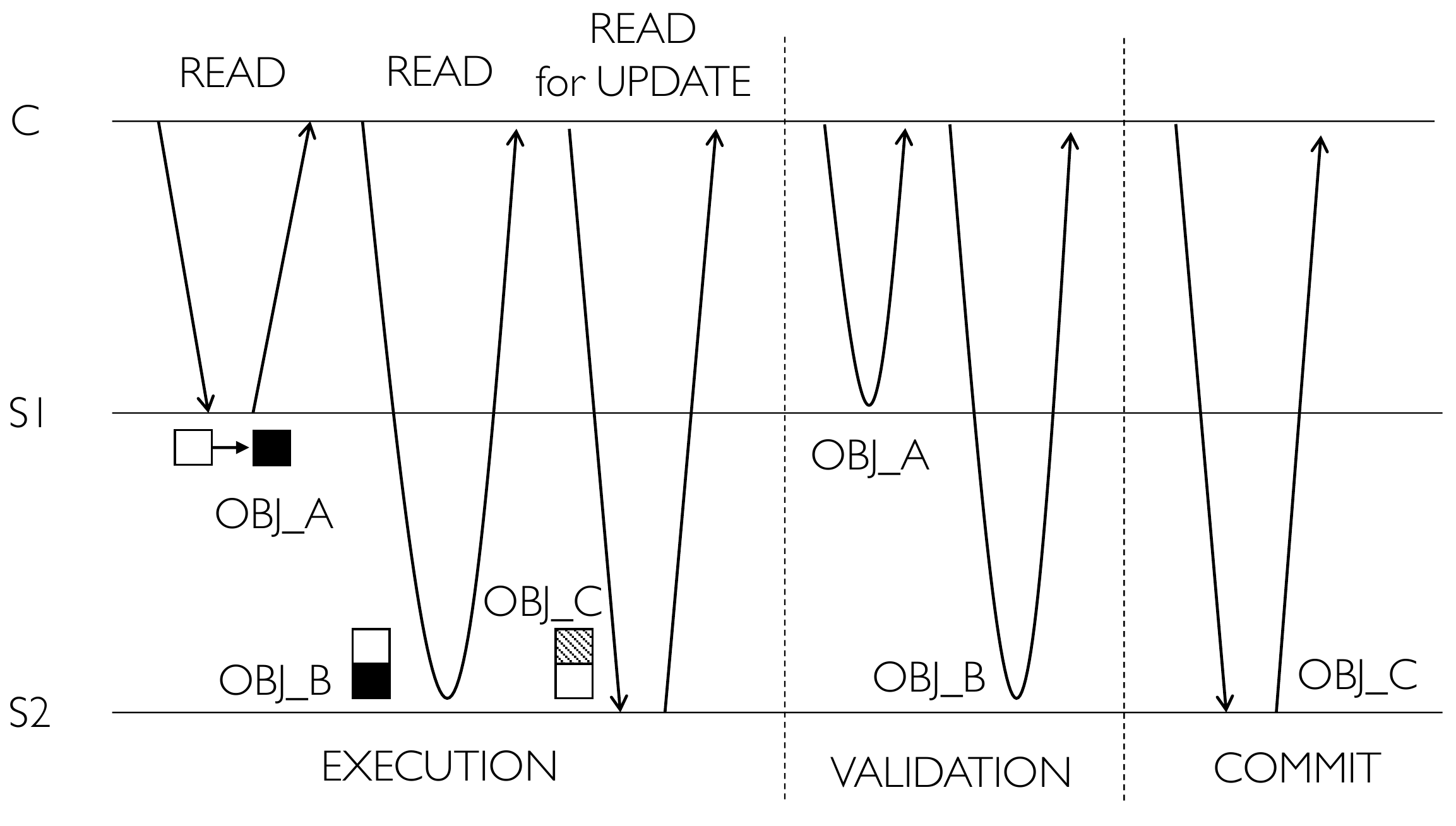}
    \caption{Storm execution phase and the commit protocol. Reads and updates performed using both remote reads and RPCs. Validation is performed using remote reads and commit using RPCs. Storm falls back to RPC if remote reads turn out to be expensive (first read on the figure). }
    \label{fig:protocol}
\end{figure}

\system is capable of executing serializable transactions efficiently.
\system implements a typical variation of the two-phase commit protocol
that is optimized for RDMA; throughout the execution phase, objects are copied into the local memory and written. Before committing, \system validates that the read set has not been modified by a concurrent transaction. This is done using remote reads, as \system keeps track of the remote offsets of each individual object in the read set. Finally, \system uses write-based RPCs to update the objects from the write set and unlock them. 
Figure~\ref{fig:protocol} illustrates how \system mixes remote reads 
and RPCs within the execution phase of a single transaction. \system
uses optimistic concurrency control~\cite{occ-TODS81}, but locks 
the objects that the transaction intends to write in the execution phase. Alternatively, the protocol could optimistically read for update (without locking) and lock in the commit phase, before the validation, as implemented in FaRM~\cite{dragojevic:farm}.

%In the execution phase, Storm utilizes both reads
%and RPCs for best throughput. Which primitive is the most
%efficient depends on the application data layout; if too many %round-trips are required to perform a lookup, Storm falls back to %RPCs. \yizhou{1.how many is too many? 2. Is this switching %completely transparent to users?}
%In this paper, we show how this can be benefitial in %high-performance RDMA storage systems such as Storm.

%Since our implementation of RPC requires polling the message buffers, \system aims to use %remote reads whenever possible, which require polling a constant number of buffers. To be %able to use remote reads for lookups, it is necessary to minimize the number of required %round-trips. We find that reducing the bucket size in a hash table leads to fewer %round-trips and more efficient remote read lookups.

\subsection{Example remote data structure: hash table}
We use a hash table as a classical remote data structure example. In addition to being remote, this hash table is also distributed, but this does not affect Storm's design in any way. We modified the MICA hash table~\cite{micaht} to accommodate for zero-copy transfers and extended it with handlers from Table~\ref{table:apicb}. Zero-copy is achieved through inlining of the required metadata, including: key, lock and version. The \textit{rpc\_handler} is compatible with Storm transactions and implements lookups, lock acquisition, updates, inserts and deletes. To lookup remote items, the clients call into \textit{lookup\_start} to get the address based on hash. The MICA hash table allows us to change buffer allocation and specify the bucket size, which we leverage to reduce hash collisions. 

Besides hash tables, Storm allows the user to implement other types of basic data structures, such as queues and stacks, trees, etc. The metadata that is cached may differ across data structures; for example, for queues the head and tail pointers may be cached on the client side. For trees, the clients could cache higher levels of the tree to improve traversals. 

\subsection{Concurrency}
Asynchronous scheduling of remote reads and RPCs is a difficult task. One could use callback 
continuations to pipeline multiple remote operations concurrently. Even though this approach has
low overhead, the preferred method has been to use user-level threads (i.e., coroutines)~\cite{kalia:fasst,hybrid-osdi18}. Storm leverages coroutines to provide  concurrency within each individual thread, while reducing the complexity of building applications
on top of the Storm TX API; from the developer's perspective all 
operations appear blocking. 

%This model is illustrated in Algorithm~\ref{alg1} with \textit{yield} function calls.

\begin{table*}[t!] 
\footnotesize
\centering 
\caption{Different evaluation platforms used in this work}
\label{table:evalplatforms}
\renewcommand{\arraystretch}{1.4} 
\begin{tabular}{|m{2.2cm} | m{4.7cm}| m{4.7cm} | m{2cm} |}
\hline 
\textbf{Platform:} & \textbf{CPU/memory} & \textbf{RDMA network} & \textbf{Max. Machines}\\
%\hline
%Operating system & \todo{XX} & \todo{XX}\\
\hline 
CX3 (RoCE)  &  \multirow{3}{*}{Intel Xeon Gold 5120, 192GB DRAM} & Mellanox ConnectX-3 Pro 40Gbps & \multirow{3}{*}{2}\\ 
\cline{1-1}\cline{3-3}
CX4 (RoCE) &  &  Mellanox ConnectX-4 VPI 100Gbps &\\
\cline{1-1}\cline{3-3}
CX5 (RoCE) &  & Mellanox ConnectX-5 VPI 100Gbps &\\
\hline 
CX4 (IB) & \multirow{1}{*}{Intel Xeon E5-2660, 128GB DRAM} & \multirow{1}{*}{Mellanox ConnectX-4 IB EDR 100Gbps} & 32\\ 
%\cline{1-1}\cline{4-4}
%CX4-real (IB) &  &  & 24\\ 
\hline
\end{tabular}

\end{table*}

\section{Evaluation}
\label{sec:eval}

\subsection{Methodology}
\label{sec:methodology}

We use Infiniband EDR to evaluate key design benefits of \system and point out the downsides of the previous proposals. We first briefly explain our experimental methodology.

\paragraph{RDMA test-bed.} We deployed and evaluated \system on a 32-node Infiniband EDR (100Gbps) cluster. Each machine features a Mellanox ConnectX-4 NIC, which has similar performance characteristics to ConnectX-5.
In addition, we have access to three pairs of servers, a pair for each of the three most recent ConnectX generations (CX3, CX4, CX5), all based on RoCE. 
Table~\ref{table:evalplatforms} summarizes our test-beds. In this section, we only focus on the Infiniband cluster because it has the largest number of nodes. 
We use a key-value micro-benchmark to evaluate workloads dominated by single-object reads (no transactions). Also, we use the Telecommunication Application Transaction Processing (TATP)~\cite{tatp-website}, which is a standard database benchmark. Both benchmarks are tested on the previously described distributed hash table. In addition to \system, we also deploy and run eRPC and our emulated and improved
version of FaRM. We were not able to deploy LITE on this cluster, 
as we were not allowed to patch the kernel on the cluster. Instead,
we ported and deployed LITE on our CX5(RoCE) servers and projected the results to our CX4(IB) platform.

%By using different RDMA NICs, we are able to present
%current technology trends in RDMA.
%In addition to different generations of RDMA NICs, we also evaluate %\system on both Ethernet-based RDMA (RoCE) and Infiniband.

%learn more about the trends and gain deeper understanding of the NIC internals. %Based on these insights, we developed \system, an efficient RDMA-based transactional storage system that will enable future applications to fully leverage the benefits of a high-performance RDMA network.
%: we built one test-bed consisting of three pairs of servers with %different generations of RoCE NICs. In addition, we have access to %one 24-machine cluster, each with CX4 NICs and connected using an %Infiniband EDR network (100Gbps). 
%We use the three pairs of RoCE servers 
%to emulate larger deployments.

\paragraph{Emulation.}With \system we are able to emulate RDMA clusters larger than 32 nodes. To achieve that, \system allocates the same amount of resources that would exist in a real environment, including connections and registered RDMA buffers. For example, each thread maintains a connection to each of its "siblings" (i.e., threads with the same local ID) on the other servers. By varying the number of QP connections and the amount of message buffers used per thread, we can accurately emulate clusters of 3-4x larger sizes. The maximum size is limited because of the amount of compute that is fixed. 

%In this paper, we will present analysis using both the 24-machine %``real'' cluster and 64-machine ``emulated'' cluster (using both IB %and RoCE). We find that the performance of Storm's RPC cannot be %faithfully emulated because the message polling overhead increases %with cluster size but the amount of clients does not. This is not %true in real envirnoments and thus we only present Storm's RPC %results for the real 24-machine cluster (\S\ref{sec:realperf}). 

%Table~\ref{table:evalplatforms} summarizes the %different evaluation platforms we use in this %work.

%We can emulate the clusters of up to 64 nodes, %beyond that, when we use RPCs (for example, %when using update transactions) the CPUs on %each of the servers becomes the bottleneck and %the network performance is no longer %representative of what will be observed on %large clusters.
%The workloads are sufficiently different to allow the reader to draw conclusions and develop greater understanding of the RDMA hardware, the \system dataplane, and RDMA-based distributed storage systems.

\paragraph{Workloads.} We use two workloads, described next. \\
%
%$\bullet$ \textit{Random reads:} This workload reads remote memory %locations randomly using a uniform distribution. 
%Reads are cache-line sized. This represents the worst-case scenario %for performance and scalability of RDMA,
%since there is no locality and the active protocol state overwhelms %the NIC caches. In addition, cache-line requests involve tiny %transfers, emphasizing any RDMA overheads.\\
%%of the RDMA interface and managing the protocol state. \\
%
$\bullet$ \textit{Key-value lookups} uses \system to look up random keys in the \system distributed hash table.
%We simply refer to this workload as \textit{Lookups}. 
Each bucket has a configurable number of slots for data. Colliding items are kept in
 a linked list when the bucket capacity is exceeded.
 %(a.k.a. overflow chain) when the bucket capacity is exceeded.
When the hash table is highly occupied, linked list traversals are needed to find a key.
Each data transfer, including the application-level and RPC-level headers, is 128 bytes in size. \\
$\bullet$ \textit{TATP} is a popular benchmark that simulates accesses to the Home Location Register database used by a mobile carrier; it is often used to compare the performance of in-memory transaction processing systems. TATP uses \system transactions to commit its operations.\\

\paragraph{Baselines.} We compare \system to three different baseline systems: (i) eRPC, which is a system based on Unreliable Datagrams (UD); (ii) FaRM, a system that leverages the Hopscotch hashtable algorithm to minimize the number of round trips; and (iii) LITE, a kernel-based RDMA system that onloads the protection functionality to improve scalability. 
eRPC does not allow for one-sided reads and is an RPC-only system. eRPC relies on UD, which is an unreliable Infiniband transport requiring onloaded congestion control and retransmissions. We emulate FaRM by configuring \system with FaRM parameters. Also, to provide a fair comparison, we do not share QPs using locks as our NICs scale better compared to the CX3, which have been used to evaluate FaRM. Finally, we improved LITE by extending it with support for asynchronous remote operations. Asynchronous operations are important for throughput-oriented applications, such as transactions.

\subsection{Performance at rack-scale}
\label{sec:realperf}
We first evaluate \system in isolation using the \textit{Key-value lookups} workload. Then, we compare \system to the previously proposed systems using the same workload, and finally we evaluate TATP running on \system.

\subsubsection{Key-value lookups}

\begin{figure}[t] %h!
    \includegraphics[width=0.47\textwidth]{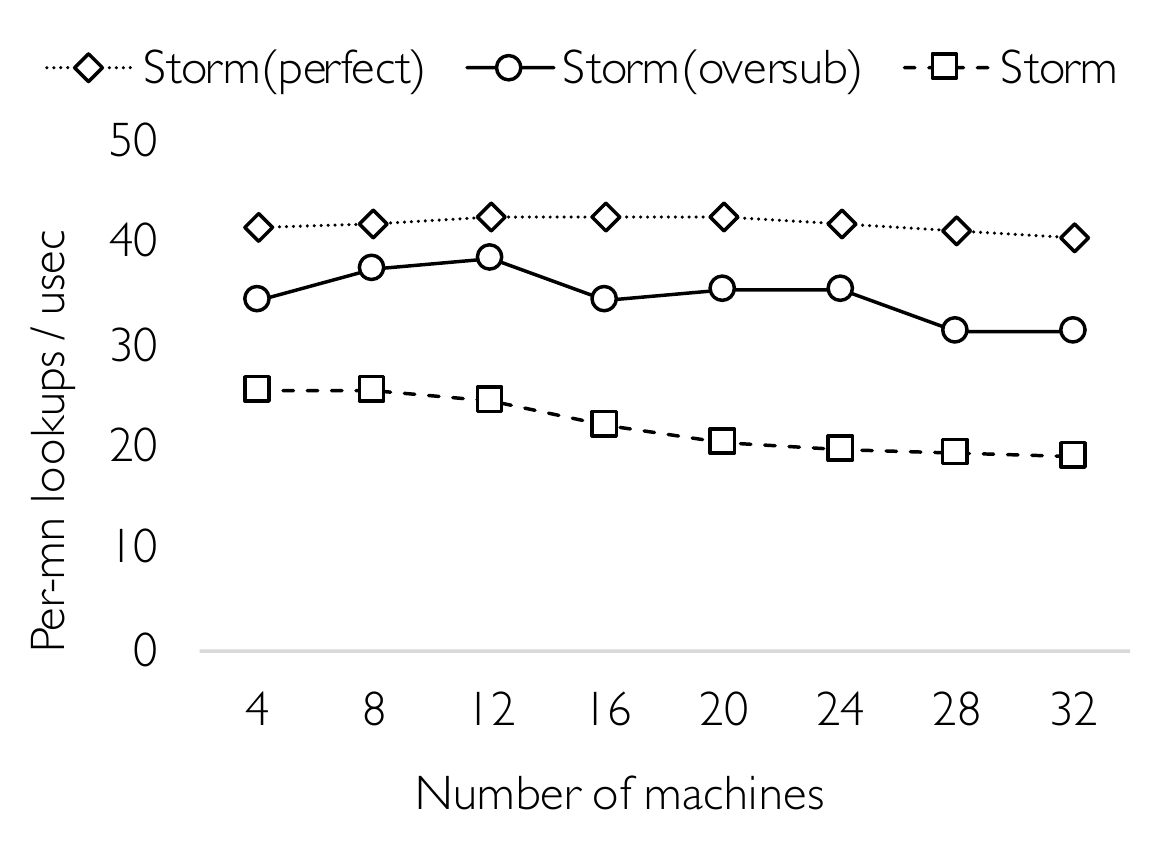}
    \caption{Comparison of different Storm configurations for a read-only key-value workload. Average per-machine throughput on the Y-axis. }
    \label{fig:commit}
\end{figure}

Figure~\ref{fig:commit} shows the performance for three different \system setups: (i) \textit{Storm} uses only RPCs to perform lookups. We observe that the throughput stabilizes with the node count; more nodes amortizes the polling overhead on the receiver. (ii) \textit{Storm(oversub)} enforces lower collision rate by allocating a larger hash table. With 32 nodes the throughput is 1.7x higher compared to \textit{Storm}. The throughput is not stable as we scale because the collision rate is not the same for different node counts, which translates to a higher or fewer number of reads followed by RPCs (\textit{one-two-sided}), impacting throughput. Finally, (iii) \textit{\system(perfect)} assumes no
RPCs on the data path. Using only remote reads in Storm is possible through a combination of memory oversubscription and caching of the addresses of pointer-linked items. At 32 nodes, \textit{\system(perfect)} outperforms \textit{\system} by 2.2x. 

\subsubsection{Key-value lookups (comparison)}
\label{sec:lookupall}

\begin{figure}[t]
    \includegraphics[width=0.47\textwidth]{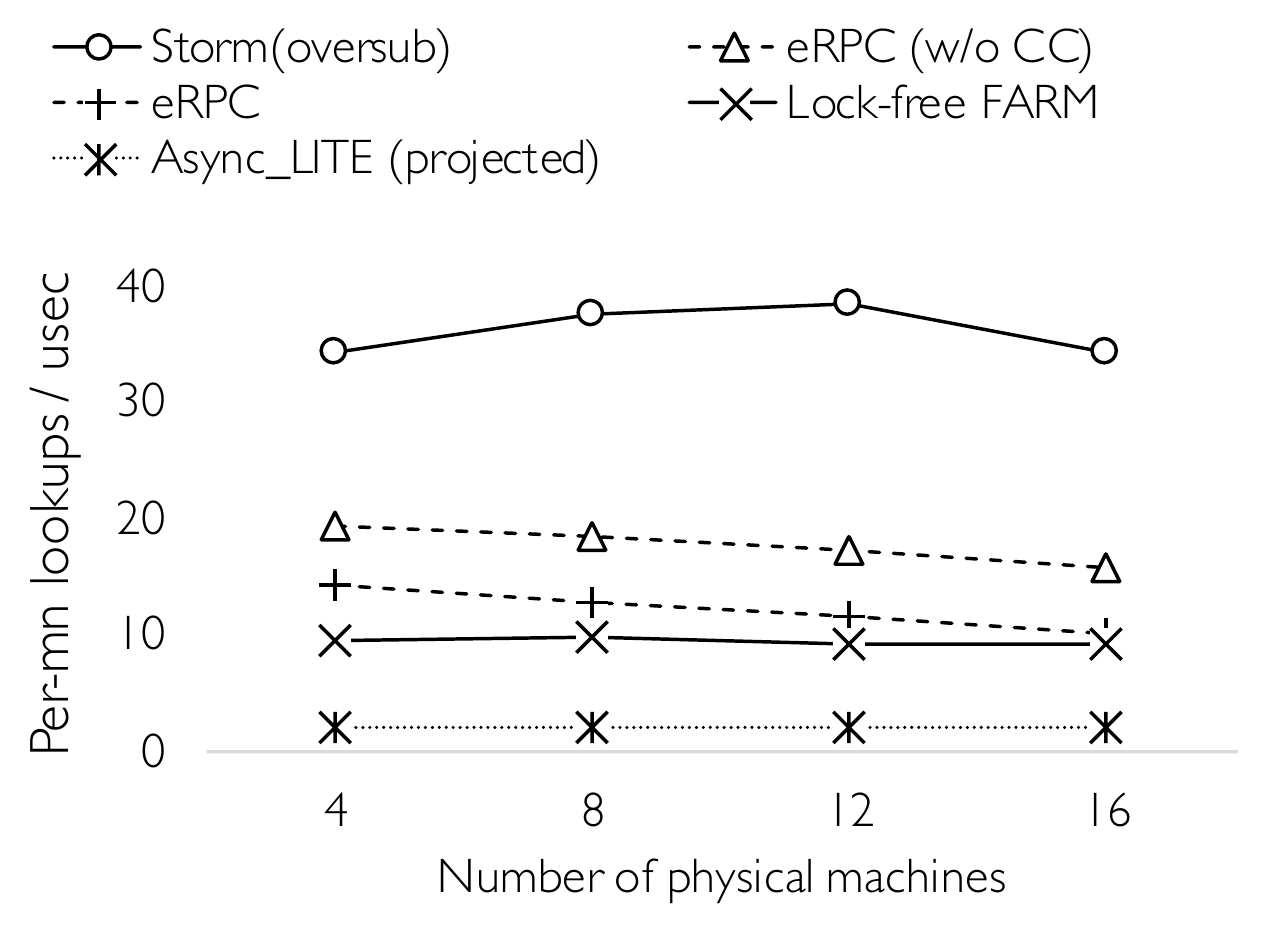}
    \caption{Comparison of \system, eRPC, FaRM, and LITE. eRPC includes two versions, with and without congestion control. We optimized both FaRM and LITE for the maximum throughput. }
    \label{fig:txbenchcomp}
\end{figure}

In this section we compare the performance of \system, eRPC, FaRM, and LITE using the \textit{Key-value lookups} workload.
Figure~\ref{fig:txbenchcomp} presents the performance of all the systems running on a real cluster with sizes varying from 4 to 16 machines. We were not able to deploy eRPC on more than 16 nodes (hence X-axis goes up to 16), as our NICs do not support sufficiently large receive queues. eRPC relies on a large-enough number of registered receive buffers to prevent receiver-side packet loss. For \system, we only plot \textit{Storm(oversub)}.
For eRPC, we study the baseline version and one without congestion control, whereas \textit{Storm(oversub)} has hardware congestion control always enabled.
For FaRM, we use our improved emulated version that does not require QP locks, unlike the original FaRM implementation~\cite{dragojevic:farm}. We emulate FaRM by configuring \system with the same parameters from the original FaRM paper~\cite{dragojevic:farm}. A key difference is that we use 128B items, which increases the bucket size in FaRM and affects throughput. Finally, we use our improved version of LITE that enables asynchronous remote reads and RPCs (\textit{Async\_LITE}).

The key takeaways are: (1) \system significantly outperforms previous systems. This gap is mainly due to Storm's ability to take advantage of fine-grain remote reads.
(2) Even though eRPC does not use a reliable transport (no connections), the throughput decreases with node count due to the increasing overhead of posting onto the receive queue. This issue can be fixed using "strided" RQ, which unfortunately is not available on our infrastructure. Strided RQ enables posting a single RQ descriptor for a set of virtually contiguous buffers. The lack of this feature also limits us to 16 nodes. This limit holds only for eRPC and not for other evaluated systems. (3) eRPC with no congestion control performs 1.53x better at 16 nodes than eRPC with application-level congestion control with bypass techniques enabled~\cite{erpc}, indicating that relying on the implicit congestion control provided by RC rather than the custom congestion control at the application level may be beneficial. The overhead of onloaded congestion control will become more problematic with decreasing network latencies and increasingly higher IOPS rates (4) FaRM with its coarse-grained reads performs worse than eRPC, suggesting that trading larger network transfers (8x) per lookup for fewer network round trips comes with performance overhead. For items smaller than 128 bytes, FaRM achieves higher throughput, as this results in smaller bucket transfers. Finally, (5) LITE performs the worst due to 
the kernel complexity. We measured the throughput on two CX5 nodes only and projected these measurements to 16 nodes. LITE is compute-bound
and does not suffer from NIC cache thrashing. Hence, we expect the throughput to be similar when running on larger clusters with CX4.

\subsubsection{TATP performance}

\begin{figure}[t]
    \includegraphics[width=0.47\textwidth]{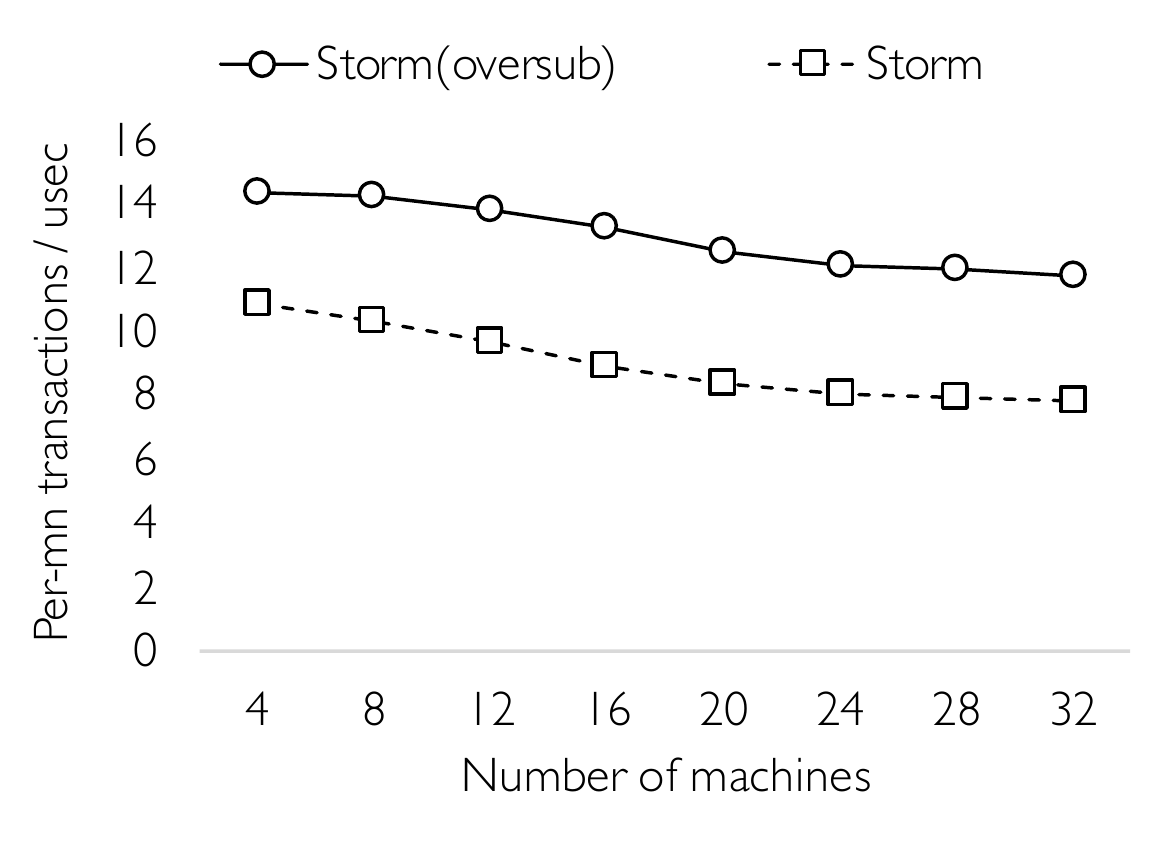}
    \caption{TATP running on Storm. Lower occupancy of TATP hash tables leads to better performance.}
    \label{fig:tatp}
\end{figure}
On Figure~\ref{fig:tatp} we study TATP for two \system configurations. Both configurations allocate the same amount of memory for the data.
The configurations are as follows: (1) \textit{\system(oversub)} uses an oversized hash table with bucket width of one, where each unsuccessful remote read lookup is followed by an RPC (\textit{one-two-sided}) to traverse the overflow chain. To perform read-for-update and commit, \system uses RPCs. The oversized hash table results in fewer collisions and the ability to successfully leverage remote reads most of the time; (2) \textit{\system} always uses RPC to execute all application requests, independent of the bucket size.

At 32 nodes, \textit{\system(oversub)} outperforms \textit{\system} by 1.49x. The TATP workload has 16\% of writes and 4\% of inserts and deletes. Writes, inserts and deletes require RPCs and thus the improvement is not as significant as in the \textit{Key-value lookups} workload. Also, with increasing node count, the throughput trend is similar to that of \textit{Storm} in the \textit{Key-value lookups} workload, and this is because of a larger fraction of RPCs.

\begin{comment}
We see that: (1) the performance of \system(RR,RPC) and \system(RPC) are comparable. This similarity is to be expected due to the comparable performance of remote reads and write-based RPCs we observed in Section~\ref{sec:lookupall}. Going from 8 to 24 machines, we observe a slightly larger slowdown for \system(RPC): ~6\% for both TATP-80 and TATP-96, and close to 0\% slowdown for \system(RR,RPC)  (2) the performance of \system(RR,RPC) and \system(RPC) is higher than that of \system7(RR,RPC). This behavior is also to be expected because linked-list traversals are much faster using RPCs than using multiple consecutive remote read operations. Nevertheless, the degradation is not as significant as one might expect, even through the occupancy of all TATP tables is ~90\%. We further investigate why this is the case and conclude that a combination of the non-uniform access pattern of TATP and non-uniform hash collisions reduce the average number of round-trips per lookup. By setting a cut-off threshold for switching from remote reads to RPCs to 7, we limit the number of round trips for long overflow chains. The non-uniform access pattern insures these highly contended buckets are not frequently targeted. By resizing the hash table and having the bucket size of one, we show it is possible to further reduce the average number of round trips and provide high throughput using remote reads. (3) Storm 7 (RPC) provides the same performance as Storm (RPC), since they both require the same number of round-trips per lookup (i.e., one) or per transaction.
\end{comment}

\subsubsection{Impact on latency}
(i) Unloaded latency: Table~\ref{tbl:lat} shows the unloaded round trip latencies of the evaluated systems on two of our CX4 platforms, Infiniband and RoCE. RoCE is generally known to have slightly higher latency compared to Infiniband. RPC latency for \system and eRPC is similar; both are optimized zero-copy implementations. FaRM requires transferring eight times larger blocks, hence higher latency. LITE has the highest latency due to the kernel overheads. (ii) Loaded latency: We have also looked at the 99th-percentile latency in the context of TATP. The tail latency keeps increasing as we add more nodes, but the system does not saturate the network and the latency is only on the order of tens of micro-seconds, far below a typical SLA (around 5ms). 
%Figure~\ref{fig:latency-tatp} shows 
%the loaded 99th-percentile latency for the two TATP configurations that we run. In both %cases the latency slightly increases going from 8 to 24 machines, as the aggregate load of %the system increases. However, 
%the system does not saturate the network and the latency is only on the order of tens of %microseconds, far below a typical SLA (around 5ms). The unloaded average latency for %\system running on the Infiniband network is 1.8 microseconds.

\subsubsection{Physical segments}
We studied the performance of \system on clusters with CX5 NICs and large persistent memories (PB scale).
With the advent of extremely dense persistent memory technologies, we anticipate that future servers will be hosting hundreds of TBs memory.
For such large memory machines, the RDMA region metadata can overwhelm the NIC caches, especially due to the MTT size.
We added support for physical segments in \system and enforce kernel-level segment registration for security reasons. We use 4KB page sizes and compare them to using \system to export application memory as a physical segment. By using 4KB pages, we emulate a PB-scale storage class memory with 1GB page size. Using physical segments vs 4KB pages leads to 32\% higher throughput.

\subsection{Discussion: beyond rack-scale}
\label{sec:emuperf}

\begin{figure}[t] %h!
    \includegraphics[width=0.47\textwidth]{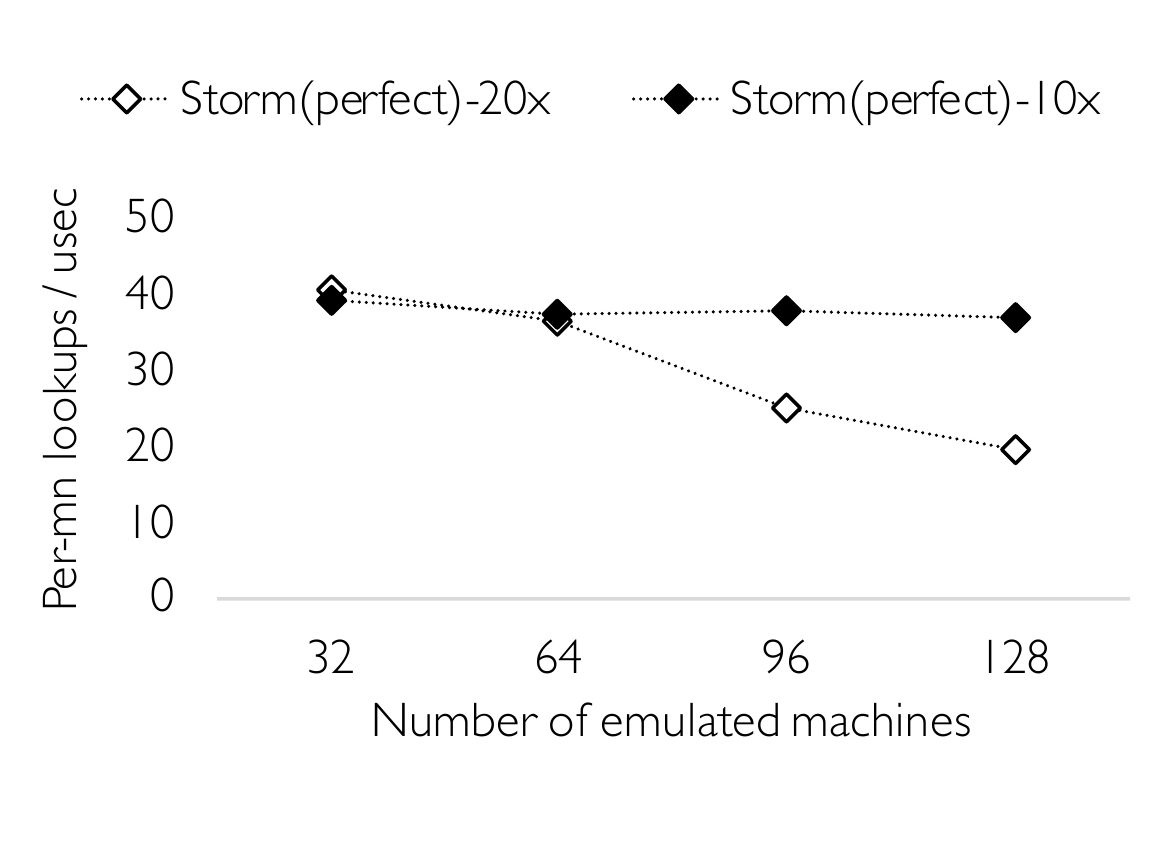}
    \caption{Emulation of larger clusters using a 32-node cluster. 128 emulated machines requires 4x more connections and RDMA buffers. Comparison of Storm(perfect) with 20 and 10 threads per machine.}
    \label{fig:emulation}
\end{figure}

In this section, we emulate larger clusters using our 32-node CX4(IB) cluster by creating additional connections and allocating additional buffers between each pair of machines~\cite{hybrid-osdi18}. Figure~\ref{fig:emulation} shows the throughput as we scale the system from 32 to 128 virtual nodes. At 96 nodes and 20 threads per (physical) node, the throughput drops by 1.57x when the NIC cache is overwhelmed with state. Most of this state consists of connections, as we minimized the amount of MTT and MPT through larger (2MB) pages and contiguous memory allocation. 

%We use all 32 nodes as the amount of compute involved also matters; with less compute there %is less NIC cache thrashing.

We observe the following: (i) Up to 64 nodes, the throughput is stable. 64 or fewer nodes is enough for most rack-scale deployments, which are most common. (ii) by reducing the number of threads to 10 per server, the throughput is stable even at 128 nodes. A smaller number of threads leads to fewer initiated connections, which minimizes the amount of transport-level state. If an application requires more than 10 threads per node, we envision a low-overhead, lock-free connection sharing mechanism that will allow Storm to scale to larger cluster sizes. We are also looking into memory management techniques for Storm to reduce memory footprint.

\begin{comment}
in which there exist two groups of threads: foreground and background. Both of them run application code and have the same QP interface to the network, but only the foreground threads have access to the physical NIC - actually register their QPs with the NIC. Background threads have their own QPs, but those are not registered with the NIC, nor they create connections with the sibling threads. Requests coming from the background threads are served by the foreground threads. Such requests are pulled periodically (e.g., when foreground queues are idle) by a single thread per queue, minimizing cache coherence traffic. 
\end{comment}
%This mechanism is currently work-in-progress.

\begin{table}[t]

\setlength{\tabcolsep}{2pt}
\footnotesize
\centering\begin{tabular}{|c|c|c|c|c|c|c|c|}
	\hline
	Platform 	& 	\system (RR) 	& 	\system (RPC) 	& 	eRPC & FaRM & LITE\\
	\hline\hline
	CX4 (IB)		&  1.8us  & 2.7us & 2.7us & 2.1us & 5.8us\\

	\cline{1-1} \cline{2-2} \cline{3-3} \cline{4-4} \cline{5-5}	\cline{6-6}	
	CX4 (RoCE)		&  2.8us  & 3.9us & 3.6us & 3us & 6.4us\\

	\cline{1-1} \cline{2-2} \cline{3-3} \cline{4-4} \cline{5-5} \cline{6-6}	
\end{tabular}
\caption{Round-trip unloaded latencies for the various baselines and \system}
\label{tbl:lat}
\end{table}

\section{Related Work}
\label{sec:related}

Other than the systems discussed in the previous sections of this paper, there is a large body of other work on RDMA-based key-value and transaction processing systems~\cite{DrTM-sosp15,Chen-eurosys16,pilaf-atc13,rfp-eurosys17,memcached-rdma,cell-atc16,kvdirect-sosp17}, distributed lock management~\cite{yoon-sigmod18,Narravula-CCgrid07}, DSM systems~\cite{grappa-atc15}, PM systems~\cite{remote-region-atc18,hotpot-socc17,octopus-atc17,crail,tavakkol2018enabling}, and resource disaggregation~\cite{infiniswap-nsdi17,zombieland-eurosys18}.
We discuss only a few of these in this section.

\paragraph{Other one-sided RDMA storage systems.}
Pilaf~\cite{pilaf-atc13} uses a self-verifying data structure to detect races and enforce synchronization. This mechanism is directly applicable to Storm.  
 NAM-DB~\cite{namdb-vldb17,namdb-vldb16} leverages multi-versioning to minimize the overhead of running distributed transactions. Storm does not focus on optimizing the commit protocol and instead focuses on improving the datapath. Crail~\cite{crail} is based on Java but provides competitive performance by cutting through the Java stack (e.g., bypasses serialization). However, Crail is better suited to data processing systems, unlike Storm, which is optimized for fine-grain one-sided transfers. 
 
 \paragraph{Hybrid RDMA systems.}
 Cell~\cite{cell-atc16} uses one-sided reads and two-sided RPC in a distributed B-tree store. \system is more general-purpose and can support various data structures. Also, Cell uses send/receive to implement RPC, which we show is sub-optimal.
 RTX~\cite{hybrid-osdi18} provides
 key insights about the choice of RDMA primitive for each phase of a two-phase commit protocol using both UD and RC transports. Unlike RTX, Storm's focus is on the scalability aspect. Storm argues for using the connected transport (RC) only and taking advantage of high-throughput one-sided primitives (even for RPC). RTX validates our conclusion that one-sided operations achieve significantly higher IOPS compared to UD-based RPC for messages larger than 64 bytes and still opts to use UD for RPC due to scalability concerns. In this work, contrary to common wisdom, we show this is not necessarily a concern.

\paragraph{Unreliable Connections in Transaction Processing.} HERD~\cite{herd-rpc-atc16} is another system using the Unreliable Datagram (UD) transport, which allows a thread to use a single QP to talk with all the machines in the cluster. Our evaluation shows that this approach also has scalability limitations and limits the maximum throughput as it does not allow for one-sided operations. We leverage this fact and use reliable connection along with contiguous memory allocation to build \system, which makes the most out of the underlying RDMA network.

\paragraph{Distributed Shared Memory}
Systems emulating shared memory using RDMA~\cite{remote-region-atc18,hotpot-socc17,octopus-atc17} fault in 
remote pages into the local memory. This mechanism is very convenient from the programming perspective but incurs significant performance overheads. Storm does not provide the memory mapping functionality but exposes an intuitive API based on memory copy and RPC primitives.

%\paragraph{Kernel-based RDMA system.} Opposite to other works, LITE~\cite{lite:tsai} onload functionalities from NIC back to kernel, and enforce protection and translation inside OS. Internally, LITE uses global physical memory region~\cite{lite:tsai} and QP sharing to improve RDMA scalibility. However, LITE is designed for systems that require low-latency but not throughput-oriented transaction processing.

%\paragraph{Data Planes.} Several RDMA-based systems were proposed in the past few years to address the poor scalability of RDMA. FaRM~\cite{dragojevic:farm} is a RDMA-based distributed computing platform, which uses QP sharing and huge pages to mitigate scalability issue. Opposite to most other works, LITE~\cite{lite:tsai} onload functionalities from NIC back to kernel and enforce protection and translation inside OS. HERD~\cite{herd-rpc-atc16}, FaSST~\cite{kalia:fasst}, and eRPC~\cite{erpc} minimize QP state by using unreliable datagrams. However, our study show that new generation NIC can scale well with thousands of QPs. Besides, we also propose contiguous memory allocation to minimize RDMA region metadata to mitigate the issues remain in new NIC. 

%\paragraph{RDMA-based Storage Systems.} Similar to \system, NAM-DB~\cite{nandb-vldb17} and Pilar~\cite{pilaf-atc13} mainly leverage one-sided RDMA primitives to achieve high-throughput transaction processing. Additionally, \system uses a novel hybrid read and write based RPCs to reduce round-trips per lookup, thus improve the application throughput.
\section{Conclusion}

We make several contributions in the RDMA space: a detailed analysis of multiple generations of RDMA hardware, showing evidence that modern RDMA hardware scales well on rack-scale clusters. Second, we introduce Storm, a high-performance and transactional RDMA dataplane that uses one-sided reads and write-based RPCs. Finally, we give a detailed evaluation of Storm and compare it to FaSST/eRPC, and improved versions of FaRM and LITE. We show that due to the recent advancements in RDMA technology, one-sided operations are effective for rack-scale systems. For best performance, Storm minimizes protocol state and uses remote reads and write-based RPCs to access data.

\bibliographystyle{plain}
\bibliography{gen-abbrev,dblp,bibliography}

\end{document}